\documentclass[aps,prl,twocolumn,longbibliography]{revtex4-2}

\usepackage{amssymb}
\usepackage{amsbsy}
\usepackage{amsmath}
\usepackage{graphicx}
\usepackage{graphics}
\usepackage{setspace}
\usepackage{array}
\usepackage{color}
\usepackage{fontenc}
\usepackage{textcomp}
\usepackage{bm}
\usepackage{float}
\usepackage[bookmarks=false,linkcolor=blue,urlcolor=blue,colorlinks,citecolor=blue]{hyperref}

\DeclareMathOperator{\Tr}{Tr}

\newcommand{\vex}[1]{\bm{\mathrm{#1}}}

\newcommand{\blue}[1]{{\color{blue}{#1}}}

\newcommand{\bsub}{\begin{subequations}}
\newcommand{\esub}{\end{subequations}}

\graphicspath{{../Figures/}}

\begin{document}
\title{Topological and stacked flat bands in bilayer graphene with a superlattice potential}
\author{Sayed Ali Akbar Ghorashi$^1$}\email{sayedaliakbar.ghorashi@stonybrook.edu}
\author{Aaron Dunbrack$^1$, Ahmed Abouelkomsan$^3$
, Jiacheng Sun$^1$, Xu Du$^1$}
\author{Jennifer Cano$^{1,2}$}\email{jennifer.cano@stonybrook.edu}
\affiliation{$^1$Department of Physics and Astronomy, Stony Brook University, Stony Brook, New York 11794, USA}
\affiliation{$^2$Center for Computational Quantum Physics, Flatiron Institute, New York, New York 10010, USA}
\affiliation{$^3$ Department of Physics, Stockholm University, AlbaNova University Center, 106 91 Stockholm, Sweden}

\date{\today}

\newcommand{\be}{\begin{equation}}
\newcommand{\ee}{\end{equation}}
\newcommand{\bea}{\begin{eqnarray}}
\newcommand{\eea}{\end{eqnarray}}
\newcommand{\h}{\hspace{0.30 cm}}
\newcommand{\vs}{\vspace{0.30 cm}}
\newcommand{\n}{\nonumber}

\begin{abstract}
We show that bilayer graphene in the presence of a 2D superlattice potential provides a highly tunable setup that can realize a variety of flat band phenomena.
We focus on two regimes: (i) topological flat bands with non-zero Chern numbers, $C$, including bands with higher Chern numbers $|C|>1$; and (ii) an unprecedented phase consisting of a stack of nearly perfect flat bands with $C=0$. For realistic values of the potential and superlattice periodicity, this stack can span nearly 100 meV, encompassing nearly all of the low-energy spectrum. We further show that in the topological regime, the topological flat band has a favorable band geometry for realizing a fractional Chern insulator (FCI) and use exact diagonalization to show that the FCI is in fact the ground state at 1/3 filling.
Our results provide a realistic guide for future experiments to realize a new platform for flat band phenomena.
\end{abstract}
\maketitle

\blue{\emph{Introduction}}.---Moir\'e heterostructures have attracted tremendous interest in recent year, exhibiting a wide variety of phases driven by electron correlations,
including superconductivity \cite{cao2018unconventional,yankowitz2019tuning,lu2019superconductors}, Chern insulators \cite{serlin2020intrinsic,nuckolls2020strongly,lu2019superconductors,chen2020tunable,xie2021fractional}, Mott insulators \cite{cao2018correlated,xu2020correlated,tang2020simulation}, and Wigner crystals \cite{regan2020mott}.
Underlying the emergence of these phenomena are flat bands.
While flat bands were theoretically predicted in twisted bilayer graphene (TBLG) over a decade ago \cite{bistritzer2011moire,morell2010flat}, seminal experiments \cite{cao2018unconventional,cao2018correlated} showing correlated insulators and superconductivity in magic-angle TBLG ignited a search for flat bands in a variety of systems.
In quick succession, new twisted graphene heterostructures entered the scene, such as twisted trilayer and double bilayer graphene \cite{liu2020tunable,he2021symmetry,burg2019correlated, shen2020correlated, cao2020tunable1, park2021tunable,xu2021tunable, chen2021electrically, hao2021electric}.
Twisted heterostructures beyond graphene include transition metal dichalcogenides \cite{regan2020mott,xu2020correlated,PhysRevLett.122.086402,wang2020correlated,PhysRevLett.121.026402,tang2020simulation,devakul2021magic,zang2021hartree,bi2021excitonic,wang2021staggered,zang2021dynamical,wietek2022tunable,xian2021realization,Klebl_2022},
magnets \cite{hejazi2020noncollinear,twistedbilayerCrI3,song2021direct}, nodal superconductors \cite{volkov2020magic,can2021high,zhao2021emergent}, and topological surface states \cite{JenTISL,wang2021moire,AaronJenTISL,JenDaniele}.

However, while twisted heterostructures realize a variety of correlated phases on demand, they are not a panacea.
Twist angle introduces disorder in the form of inhomogenous angle and strain.
Devices are further complicated by domain formation, lattice relaxation, and the impact of the substrate.
Combined, these factors severely hinder sample reproducibility \cite{lau2022reproducibility}.


Thus, it is desirable to explore alternative platforms to realize flat bands and moir\'e physics.
From an electronic structure perspective, the main effect of a twisted moir\'e heterostructure is to introduce both interlayer tunneling and interlayer potentials on the moir\'e length scale.
The latter can be reproduced by imposing a spatially modulated electric field, which has already been realized on monolayer graphene by inserting a patterned dielectric superlattice between the gate and the sample, with a periodicity as small as 35nm \cite{forsythe2018band}.
Such a gate-defined superlattice potential also offers control over the superlattice symmetry and geometry.

\begin{figure}[tb!]
    \centering
    \includegraphics[width=0.43\textwidth]{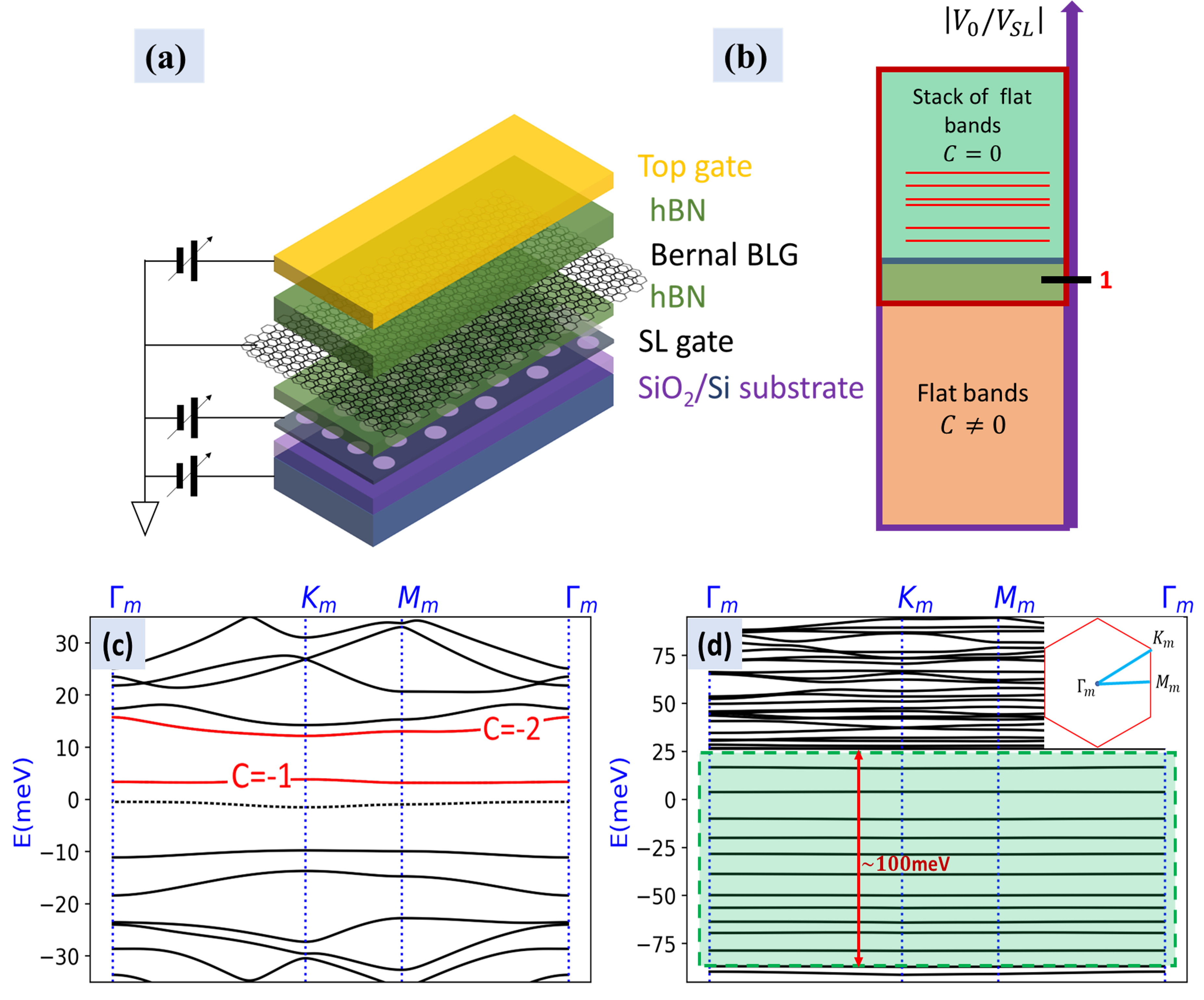}
    \caption{(a) The proposed experimental setup allows for a tunable displacement field, $V_0$, and spatially varying superlattice potential, $V_{SL}$. (b) Schematic phase diagram showing stacked and topological flat bands. (c) Energy spectrum of Eq.~\eqref{BLGSL} in the weak field limit, exhibiting a flat Chern band (red, $C=-1$) for representative parameters $V_{SL}=10,\,V_0=-5$ meV. A higher Chern band also appears (red, $C=-2$). Dotted lines indicate the low-energy bands of BLG in the limit $V_{SL}=0$.
    (d) The strong field limit exhibits a stack of flat bands (shaded green area) for representative parameters $V_{SL}=50,\,V_0=-70$ meV; mBZ in inset.}
    \label{fig:adpic}
\end{figure}

We introduce Bernal stacked bilayer graphene (BLG) in the presence of a superlattice (SL) potential as a tunable and realistic platform to realize topological flat bands.
We envision the experimental set-up depicted in Fig.~\ref{fig:adpic}(a), where BLG is subject to a spatially varying gate that creates the superlattice, denoted the SL gate, in addition to spatially constant top and bottom gates.
We find two distinct regimes of flat bands with possible sub-meV bandwidth, shown schematically in Fig.~\ref{fig:adpic}(b), with their corresponding band structures in Fig.~\ref{fig:adpic}(c,d). In the first regime, the flat bands possess a non-trivial (valley) Chern number, $C\neq 0$.
Importantly, the flat Chern bands have a near-ideal band geometry \cite{jackson_geometric_2015,roy2014band,PhysRevB.85.241308} and realize a fractional Chern insulator (FCI) at $1/3$ filling, as we will demonstrate below.
Moreover, unlike TBG, our system also realizes sought-after bands with a higher Chern number, $|C|>1$, which could give rise to exotic FCIs without Landau level analogues \cite{PhysRevLett.109.186805,PhysRevB.86.241112,PhysRevB.86.201101,sterdyniak2013series,wu2013bloch,moeller2015fractional,wu2015fractional,behrmann2016model,andrews2018stability,andrews2018stability,wang2022hierarchy}.

The second regime describes a stack of perfect flat bands with $C=0$, but non-zero Berry curvature.
This regime does not require fine-tuning, a situation unprecedented in TBLG.
Remarkably, for a reasonably strong superlattice potential, this stack can span $\sim \!\! 100$ meV, covering most of the relevant energy spectrum.
In both regimes, we study the role of the superlattice potential period, geometry, and relative potential on each layer, providing a practical guide for experimental realization of in-situ gate-tunable  flat band phenomena.

\blue{\emph{Model}}.---We model biased Bernal BLG in the presence of a superlattice potential by the Hamiltonian
\begin{equation}\label{BLGSL}
    \hat{H} =  \hat{H}_{BLG} + \hat{H}_{V_0} + \hat{H}_{SL},
\end{equation}
where the three terms describe the Hamiltonian of bilayer graphene, an applied displacement field, and a spatially varying superlattice potential, respectively.
Each term is of the form $\hat{H}_i =  \int d^2 \mathbf{r} H_i (\mathbf{r}) \hat{\Psi}^\dagger(\mathbf{r}) \hat{\Psi}(\mathbf{r})$,
where $\hat{\Psi}(\mathbf{r})$ is the electron annihilation operator at position $\mathbf{r}$, which has implicit layer, sublattice, and valley indices. We now describe each term in detail:
\begin{equation}
    H_{BLG}(\mathbf{r})  = \hbar v \tau^0 (-i\chi \partial_x \sigma^1 -i\partial_y \sigma^2)+\frac{t}{2}(\tau^1\sigma^1-\tau^2\sigma^2)
    \label{HBLG}
\end{equation}
describes biased Bernal BLG,
with $\chi=\pm$ the valley index and $t$ the interlayer coupling; Pauli matrices
$\tau$ and $\sigma$ correspond to the layer and sublattice spaces.
A displacement field $V_0$ is included via:
\begin{equation}
    H_{V_0}(\mathbf{r})  =
    V_0 \tau^3\sigma^0
    \label{HV0}
\end{equation}
Finally, the spatially modulated superlattice potential is described by
\begin{align}
    H_{SL}(\mathbf{r})=& \frac{V_{SL}}{2}\left[(\tau^0 \! +\! \tau^3)+\alpha(\tau^0 \! -\! \tau^3)\right]\sigma^0\sum_{n} \cos(\vex{Q}_n \cdot \bf{r}),
    \label{HSL}
\end{align}
where $V_{SL}$ is the strength of the superlattice potential and the set of $\vex{Q}_n$ are its wave vectors.
We specialize to the case of a triangular superlattice potential with $\vex{Q}_n=Q(\cos(2n\pi/6),\sin(2n\pi/6))$, $n=1,\ldots,6$, which define the ``mini Brillion zone'' (mBZ) by $\Gamma_m = (0,0)$, $M_m = \frac{1}{2}\vex{Q}_0$, and $K_m = \frac{1}{3}\left( \vex{Q}_0 + \vex{Q}_1\right)$, as shown in the inset to Fig.~\ref{fig:adpic}(d).
Note that $\Gamma_m$ corresponds to the original $K$ point of BLG.
The parameter $\alpha$ is the ratio of the superlattice potential felt on one layer relative to the other; the asymmetry between the layers results from the experimental set-up (see Fig.~\ref{fig:adpic}(a)) where the superlattice gate is applied to only one side of BLG.
To be concrete and realistic, in the calculations that follow we take the periodicity of the superlattice to be $L = 50$ nm and the ratio of the potential in each layer to be $\alpha=0.3$ \cite{rokni2017layer}.
We discuss the effects and physical implications of varying $L$ and $\alpha$ at the end of the manuscript and in the Supplemental Material (SM) \cite{sm}.

In the proposed setup shown in Fig.~\ref{fig:adpic}(a), $V_{SL}$, $V_0$ and the overall electron density can be tuned independently through the three gates.
Thus, there is a vast phase space in which to explore both regimes depicted in Fig.~\ref{fig:adpic}(b).

\blue{\emph{Flat Chern bands in the weak field limit}}.---
In the absence of a superlattice potential ($V_{SL}=0$), the gate bias $V_0$ opens a gap at the Dirac points (labelled by $\Gamma_m$ in the mBZ), which flattens the dispersion at the mBZ center. This gap has been well studied experimentally \cite{ohta2006controlling} and theoretically \cite{mccann2006asymmetry,mccann2006landau,castro2007biased,min2007ab}.
Since the gap has opposite signs in the two valleys, the result is a valley Chern insulator, which exhibits the valley Hall effect \cite{xiao2007valley,PhysRevB.84.075418,yao2008valley,xiao2012coupled,gorbachev2014detecting,mak2014valley}.

Starting from the valley Chern insulator, the role of the superlattice potential $V_{SL}$ is to open gaps at the boundaries of the mBZ, creating an isolated Chern band whose bandwidth is given approximately by the difference between the energy at the mBZ boundary and the gaps opened by $V_0$ and $V_{SL}$.
Since the size of the mBZ scales like $1/L$, appropriate choices of $L$, $V_0$, and $V_{SL}$ will yield a nearly flat Chern band gapped from the rest of the spectrum.

We verify this argument by a numerical calculation of the spectrum of Eq.~(\ref{BLGSL}) for a superlattice strength $V_{SL}=10 $ meV and displacement potential $V_0=-5$ meV.
The result is shown in Fig.~\ref{fig:adpic}(c): the lowest energy conduction band possesses $C=-1$ and has a very small bandwidth, only $0.66$ meV.
The indirect gaps above and below the flat band are $8.3$ meV and $3.6$ meV, respectively.
Our calculation also reveals an unexpected feature in the band structure: the next band above the gap is also topological, with a higher Chern number $C=-2$, although it is less flat.
Flat bands with higher Chern number $|C|>1$ are intriguing and sought after because they have no analogue in Landau levels and can realize exotic phases at fractional filling \cite{PhysRevLett.109.186805,PhysRevB.86.241112,PhysRevB.86.201101,sterdyniak2013series,wu2013bloch,moeller2015fractional,wu2015fractional,behrmann2016model,andrews2018stability,andrews2018stability,wang2022hierarchy}. We emphasize that while the flatness is achieved by optimizing the superlattice potential strength, the appearance of Chern bands does not require fine-tuning.

The triangular superlattice potential, unlike the square geometry,  induces a particle-hole asymmetry in the spectrum, as is evident from Fig.~\ref{fig:adpic}(c,d). However, for a weak superlattice potential $V_{SL}$, the two lowest energy bands (dotted band and red band in Fig.~\ref{fig:adpic}(c)) enjoy an approximate particle-hole symmetry. As $V_0$ and $V_{SL}$ are turned up, multiple band inversions result in a vast and complex space of band structures.
In the following, we explore this phase space to determine the effect of the superlattice potential on the bandwidth and topology of BLG.


\begin{figure}[tb!]
    \centering
    \includegraphics[width=0.49\textwidth]{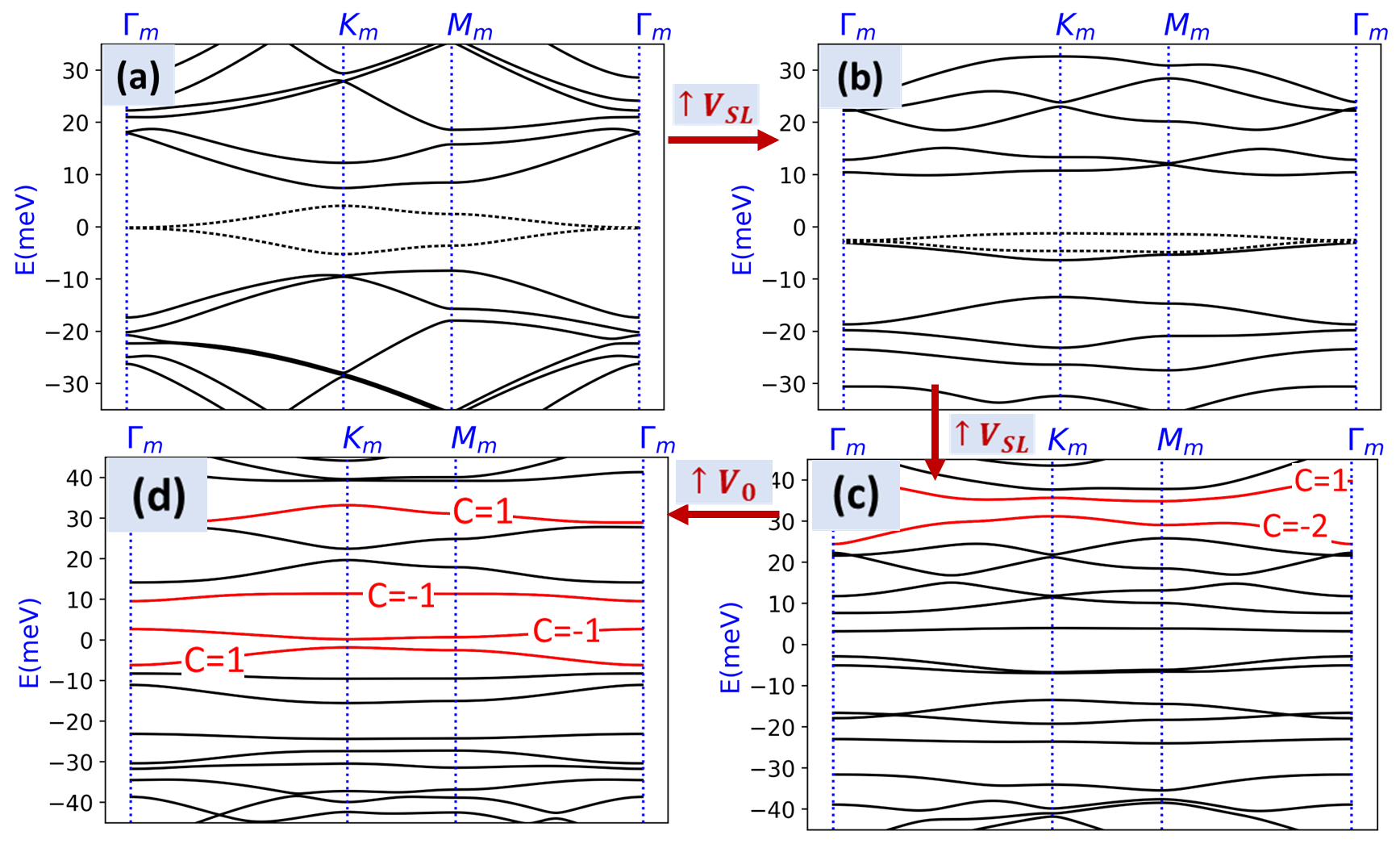}
    \caption{Band evolution of \eqref{BLGSL} upon turning up $V_{SL}$.
    (a) At zero displacement field ($V_{SL}=5,\,V_0=0$ meV), the combined Chern number of the dotted bands is $C=-1$. Turning up $V_{SL}$ first yields
    (b) a triple degenerate point ($V_{SL}=16,\,V_0=0$ meV) and then a trivial gap, shown in (c) for $V_{SL}=20,\,V_0=0$ meV.
    (d) Turning on $V_0$ from (c) opens topological gaps ($V_{SL}=20,\,V_0=24$ meV).
    Red lines show topological bands with Chern numbers indicated.}
    \label{fig:V0_0}
\end{figure}
\blue{\emph{Flat band engineering with a superlattice potential}}.---
Instead of starting from the valley Chern insulator described above, we now consider $V_{0}=0$ and slowly turn on $V_{SL}$ (Fig.~\ref{fig:V0_0}).
$V_{SL}$ opens gaps at the mBZ boundary, resulting in two low-energy bands (dotted lines), which correspond to the low-energy bands of BLG in the absence of $V_{SL}$, that detach from the rest of the bands but remain gapless at $\Gamma_m$ in the absence of $V_0$.
These two bands have a combined Chern number $C=-1$: consistent with our previous argument, turning on small $V_0$ will open the gap at $\Gamma_m$ and split them into a trivial and a Chern band (Fig.~\ref{fig:adpic}(c)).
Keeping $V_0=0$ and continuing to turn up $V_{SL}$, the two low-energy bands remain gapless up to a critical value of $V_{SL}=16$ meV where they merge with a third band to form a triple degeneracy at $\Gamma_m$, shown in Fig.~\ref{fig:V0_0}(b).
Further increasing $V_{SL}$, a small gap opens at $\Gamma_m$ between the two original bands. Though none of the low-energy bands possess $C\neq0$ (see Fig.~\ref{fig:V0_0}(c)),
relatively flat topological bands emerge at higher energies. Surprisingly, higher Chern number bands appear again, e.g., $C=-2$ in Fig.~\ref{fig:V0_0}(c).
Turning up $V_0$ from Fig.~\ref{fig:V0_0}(c) yields several Chern bands with $|C|=1$, both at the Fermi level and at higher energies, as shown in
Fig.~\ref{fig:V0_0}(d).
Summarizing, a triangular superlattice potential, $V_{SL}$, not only opens a gap at the mBZ boundary but also can induce flat topological bands, including those with Chern numbers $|C|>1$. This can occur even in the absence of the displacement field, $V_0$.


%
\begin{figure}[b!]
    \centering
    \includegraphics[width=0.5\textwidth]{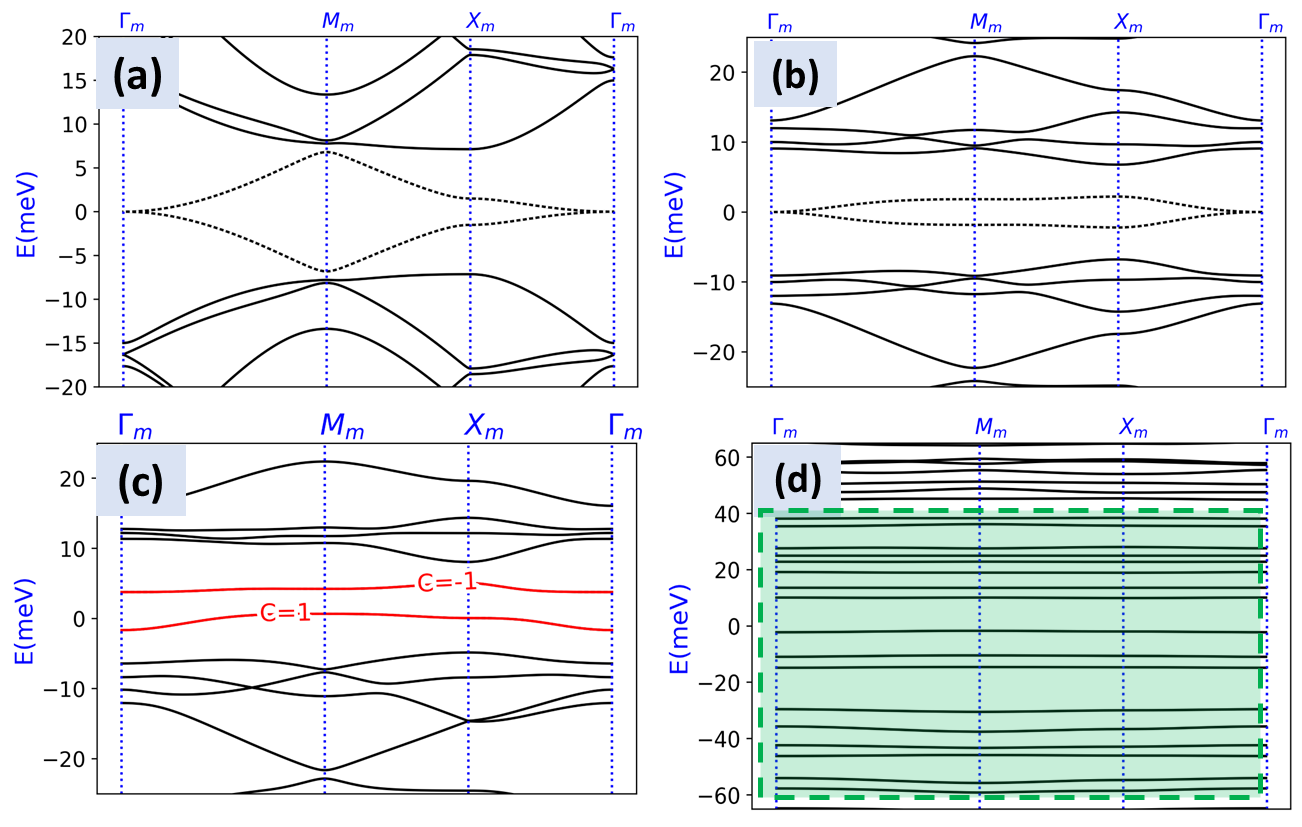}
    \caption{Band structures for a square superlattice potential. Band structure of Eq.~\eqref{BLGSL} with (a) $V_{SL}=5,\,V_0=0$ meV; (b) $V_{SL}=30,\,V_0=0$ meV; (c) $V_{SL}=30,\,V_0=-5$ meV shows red Chern bands; and (d) $V_{SL}=45,\,V_0=-65$ meV shows stack of flat band.}
    \label{fig:SQlattice}
\end{figure}

\blue{\emph{Stack of flat bands}}.---
As topological gaps open away from the original low energy bands of BLG, leading to Chern bands at higher energies, turning up $V_0$ causes multiple phase transitions and induces a larger gap between the conduction and valence bands.
Ultimately, a new regime appears, exhibiting a \emph{stack of flatbands}, indicated by the shaded green region in Fig.~\ref{fig:adpic}(d).
While these almost perfectly flat bands have vanishing Chern number ($C=0$), they have non-vanishing Berry curvature.
Thus, the electrons are not completely localized in real space.
Furthermore, the small bandwidth of the flat bands makes them highly susceptible to the Coulomb interaction, creating a quantum simulator for correlation-driven physics, similar to flat bands in moir\'e heterostructures \cite{kennes2021moire} but with complete tunability over symmetry and geometry via to the superlattice gate.

The flat band regime can be realized for both signs of $V_0$, although the spectrum is asymmetric under $V_{0}\rightarrow -V_{0}$ from the asymmetry of the experimental set-up (Fig.~\ref{fig:adpic}(a)) where the superlattice potential is applied to only one side of the heterostructure. The asymmetry enters Eq.~(\ref{BLGSL}) by setting $|\alpha| \neq 1$. Empirically, when $V_0$ and $V_{SL}$ have opposite signs, a weaker $V_0$ is required to realize the stack of flat bands (see \cite{sm} for details).

At stronger fields, and keeping $V_0 > V_{SL}$, the stack of flat bands becomes dramatically wider.
This is illustrated in Fig.~\ref{fig:adpic}(d) with $V_{SL}=50,\,V_0=-70$ meV.
The stack of flat bands span nearly $\sim \!\! 100$ meV, without fine-tuning $V_0$ or $V_{SL}$.
A phase with flat bands spanning a wide energy range has not been observed in moir\'e materials and is in sharp contrast to TBLG, which requires the twist angle be tuned very near the magic angle to realize a single set of isolated flat bands near charge neutrality.
\begin{figure}[t!]
    \centering
    \includegraphics[width=0.49\textwidth]{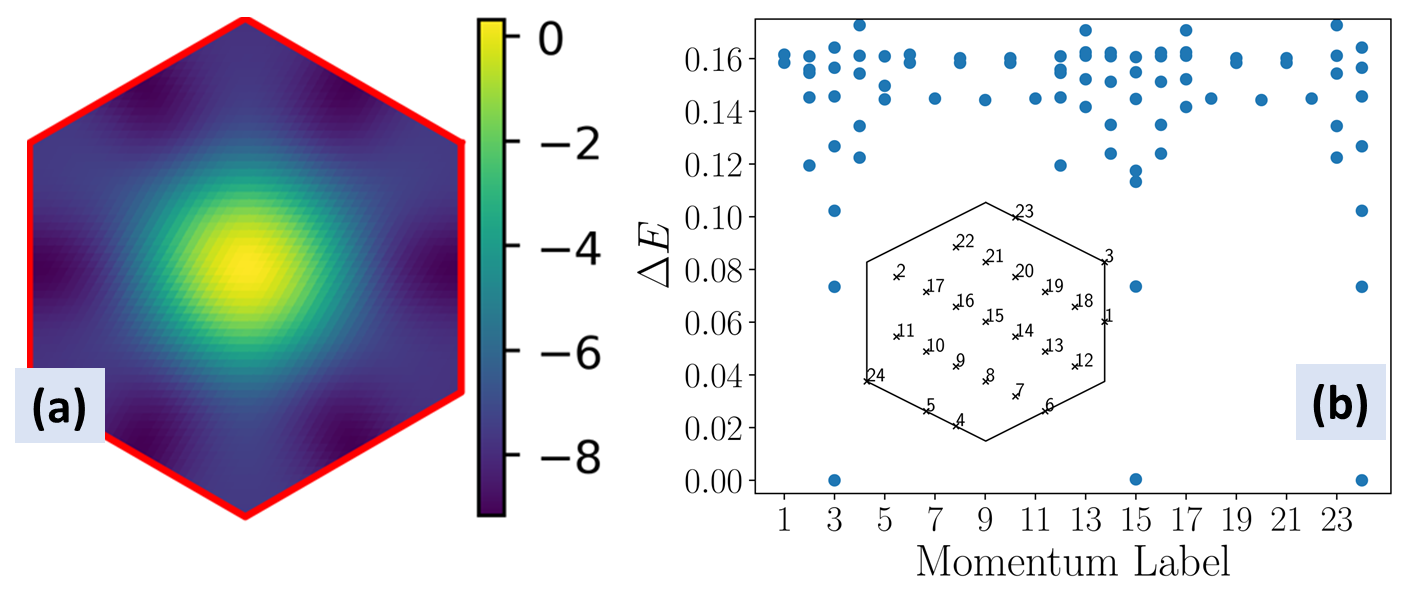}
    \caption{(a) The Berry curvature of the topological flat band labelled by $C=-1$ in Fig. 1(c) in the first mBZ (inset to Fig. 1(d)). (b) The many-body spectrum (defined relative to the lowest energy value) obtained from momentum space exact diagonalization including a dual-gated Coulomb potential projected onto the $C = -1$ band in Fig.~\ref{fig:adpic}(c) at filling $\nu = 1/3$. The inset shows the geometry of the finite cluster used.}
    \label{fig:FCI}
\end{figure}

\blue{\emph{Role of superlattice period and harmonics}}.---
The superlattice period $L$ provides another experimentally accessible tuning knob.
While Fig.~\ref{fig:V0_0} was computed with $L=50$ nm, the same phenomena appear for any value of $L$ \cite{sm}.
Optimizing the value of $L$ in an experiment must balance two considerations: (i) larger $L$ leads to flatter bands at smaller applied fields, making it easier to achieve correlation-driven physics; versus (ii) large $L$ corresponds to a large supercell more susceptible to disorder.
Further considerations depend on the precise platform.

\blue{\emph{Role of $\alpha$}}.---
The charge distribution of multilayer graphene in a superlattice potential is a complicated problem due to electron screening.
We chose $\alpha = 0.3$ following Ref.~\cite{rokni2017layer}.
To ensure our conclusions are not sensitive to this choice, we explored several other values of $\alpha$.
While changing $\alpha$ qualitatively changes the band structure,
the main features discussed in this work, i.e, the generation of topological flat bands and the stack of perfect flat bands, remain intact.
Band structures for different values of $\alpha$ are shown in \cite{sm}, including the special limits $\alpha = 1$ and $\alpha = -1$ studied in Refs.~\cite{BLGSLArunPRL,BLGSLArunPRB}.
These values of $\alpha$ are not achievable by the experimental set-up in Fig.~\ref{fig:adpic}(a), but could be realized by applying two spatially modulated fields symmetrically to the top and bottom layers of BLG, with same or opposite sign.

\blue{\emph{Lattice geometry}}.---
To investigate the role of superlattice geometry, we show that a square lattice potential yields the same phases achieved with a triangular lattice (Fig.~\ref{fig:adpic}(b)),
with qualitative differences. Fig.~\ref{fig:SQlattice}(a-c) shows that, unlike the triangular geometry, in the absence of a displacement field the spectrum remains particle-hole symmetric and gapless for all $V_{SL}$.
The square lattice is less favorable for realizing isolated topological flat bands in the weak potential limit, but tends to require a relatively weaker displacement field $V_0$ to realize stacks of flat bands (Fig.~\ref{fig:SQlattice}(d)). The two geometries and their symmetries are described in detail in \cite{sm}.

\blue{\emph{Connection to previous work}}.---
Our study of BLG is the first to show a spatially modulated 2D potential creates topological flat bands.
It differs from previous studies of BLG in a superlattice potential \cite{BLGSLArunPRL,BLGSLArunPRB,ramires2018electrically} in three fundamental ways:
(i) we consider a 2D superlattice;
(ii) we consider all four low-energy bands instead of only the lowest two, which is crucial to model the band structure at energies above the interlayer coupling strength; and
(iii) importantly, we consider a realistic experimental platform where the spatially modulated field is imposed on only one side of the heterostructure (see Fig.~\ref{fig:adpic}(a)).
Previous studies of a superlattice potential on monolayer graphene \cite{park2008anisotropic,PhysRevLett.101.126804,park2009making,PhysRevB.81.075438,dubey2013tunable,ponomarenko2013cloning,forsythe2018band,huber2020gate,li2021anisotropic,lu2022synergistic} and transition metal dichalcogenides \cite{shi2019gate,TMDSLexc} did not study topological flat bands.

\blue{\emph{Fractional Chern insulator}}.--- The competition between FCIs and symmetry-broken phases in topological flat bands is of intense current interest \cite{abouelkomsan2020particle,repellin2020chern,ledwith2020fractional,wang2021exact,wilhelm2021interplay,li2021spontaneous,abouelkomsan2022quantum}.
The FCI stability is impacted by both bandwidth and band geometry.
We have already demonstrated (Fig.~\ref{fig:adpic}(c)) that our platform realizes topological flat bands with sub-meV bandwidth.
We now demonstrate their near-ideal band geometry by computing the BZ averaged trace condition: $\overline{T}=\langle T(k)\rangle_{BZ}=\langle \Tr[g(k)]-|\Omega(k)|\rangle_{BZ}$, where $g(k)$ and $\Omega(k)$ are the quantum metric and Berry curvature (shown in Fig.~\ref{fig:FCI}(a)), respectively \cite{jackson_geometric_2015,roy2014band,PhysRevB.85.241308}. We find $\overline{T}\sim2.15$, which is a slight improvement over the estimate $\overline{T}= 4$ in TBG \cite{ledwith2020fractional}.

Thus, the band geometry is favorable for realizing an FCI ground state.
To verify this single-particle prediction, we perform an exact diagonalization study of the interacting problem of a long range dual-gated Coulomb potential projected onto the Chern band and neglecting its small dispersion. For small system sizes, we find the ground state to be spin and valley polarized (see SM\cite{sm}). We then compute the many body spectrum  assuming spin and valley polarization for a larger system size, shown in Fig.~\ref{fig:FCI}(b). We find clear signatures of a Laughlin-like FCI at fractional filling $\nu = 1/3$, specifically, the threefold many-body ground state degeneracy on the torus, shown in Fig.~\ref{fig:FCI}(b), as well as the expected spectral flow and state counting from entanglement spectroscopy \cite{li_entanglement_2008,bernevig_emergent_2012,sterdyniak_extracting_2011,wu_haldane_2014,regnault_entanglement_2015}, shown in the SM \cite{sm}.

\blue{\emph{Discussion}}.---We introduced BLG in the presence of a superlattice potential as an alternative and tunable platform to realize moir\'e physics, where the superlattice symmetry and geometry can be chosen on demand.
We proposed a realistic experimental design to realize two regimes of gate-tunable flatbands. The first regime exhibits topological flat bands with $C\neq 0$ and, in some instances, more exotic higher Chern numbers with $|C|>1$.
Of particular interest is a isolated $C=-1$ band with sub-meV bandwidth, whose quantum geometry is favorable for realizing an FCI ground state at fractional filling.
This single-particle prediction is verified by exact diagonalization including a screened Coulomb interaction projected into the topological flat band, which reveals a Laughlin-like ground state.
A more thorough multi-band calculation will be carried out in future work.

The second regime realizes a stack of many isolated almost perfectly flat bands with $C=0$. Again the bandwidth is $\sim 1$ meV.
Although these bands are topologically trivial, they have non-vanishing Berry curvature and may also exhibit interesting correlated phases at integer or fractional filling.
The possibility of superconducting phases analogous to the observation in TBG \cite{cao2018unconventional,yankowitz2019tuning,lu2019superconductors} are of particular interest.

Our results motivate the experimental study of BLG with a superlattice potential to achieve topological and non-topological flat bands without fine-tuning twist angle or introducing twist disorder. Recently, we have shown that similar phases may be realized in multilayer graphene in the presence of a superlattice
potential \cite{PhysRevB.107.195423}.

\emph{Acknowledgements}.--
S.A.A.G. and J.C. thank Jie Wang for technical guidance and feedback on the manuscript.
This work was supported in part by the Air Force Office of Scientific Research under Grant No. FA9550-20-1-0260 (S.A.A.G. and J.C.) and by the National Science Foundation through the
NSF MRSEC Center for Precision-Assembled Quantum Materials DMR-2011738 (A.D.) and under Grant No. DMR-1808491 and DMR2104781 (J.S. and X.D.).
The Flatiron Institute is a division of the Simons Foundation.


%


\newpage \clearpage

\onecolumngrid

\begin{center}
	{\large
Topological and stacked flat bands in bilayer graphene with a superlattice potential
	\vspace{4pt}
	\\
	SUPPLEMENTAL MATERIAL
	}
\end{center}

\section{Evolution of bandstructure varying superlattice period $L$}
\begin{figure}[H]
    \centering
    \includegraphics[width=1\textwidth]{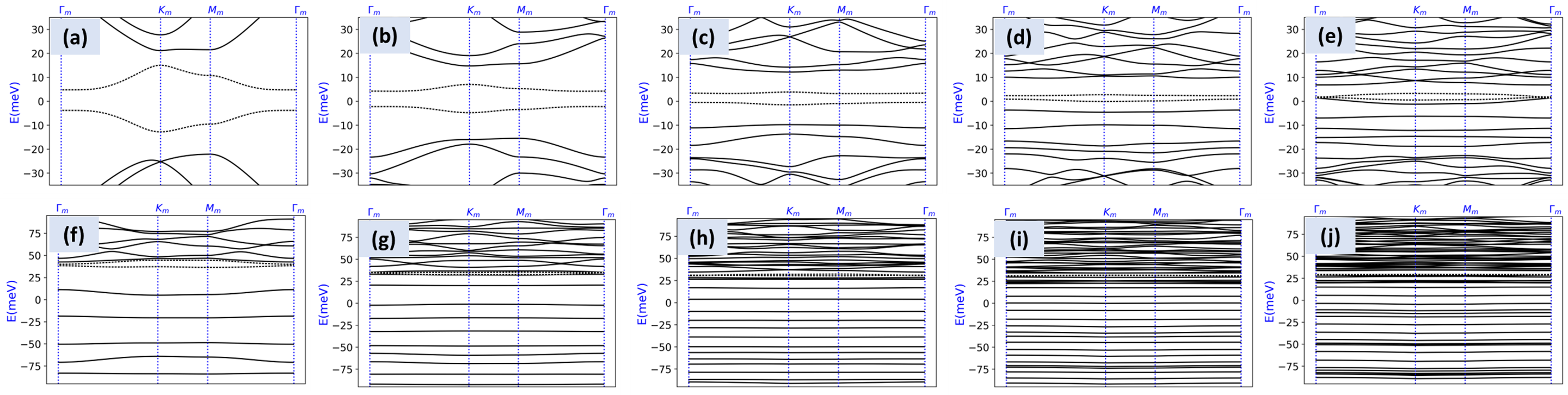}
    \caption{Evolution of the band structure of BLG in a superlattice potential in the limit of weak (first row: $V_{SL}=10,\,V_0=-5$ meV) and relatively strong (second row: $V_{SL}=50,\,V_0=-7$ meV) superlattice potential, for varying values of $L$:  (a,f) $L=30$ nm; (b,g) $L=40$ nm; (c,h) $L=50$ nm; (d,i) $L=60$ nm; (e,j) $L=70$ nm.}
    \label{fig:SMvsL}
\end{figure}

Fig.~\ref{fig:SMvsL} shows the band structure of BLG with a superlattice potential for different values of superlattice length $L$ in the limit of a weak (first row) and relatively strong (second row) potential strength.
The plots in each row are plotted with the same energy window for all values of $L$.
In Fig.~\ref{fig:SMvsL}(a-d), the upper dashed bands have a non-vanishing Chern number $C\neq 0$.
As discussed in the main text, large $L$ implies flatter bands because the mBZ is smaller.
Moreover, for larger $L$, the number of bands in a given energy window increases for the same reason.
Combined, these effects yield the stack of flat bands that appears in Fig.~\ref{fig:SMvsL}(f-j).
\section{Role of $\alpha$}
\begin{figure}[H]
    \centering
    \includegraphics[width=1\textwidth]{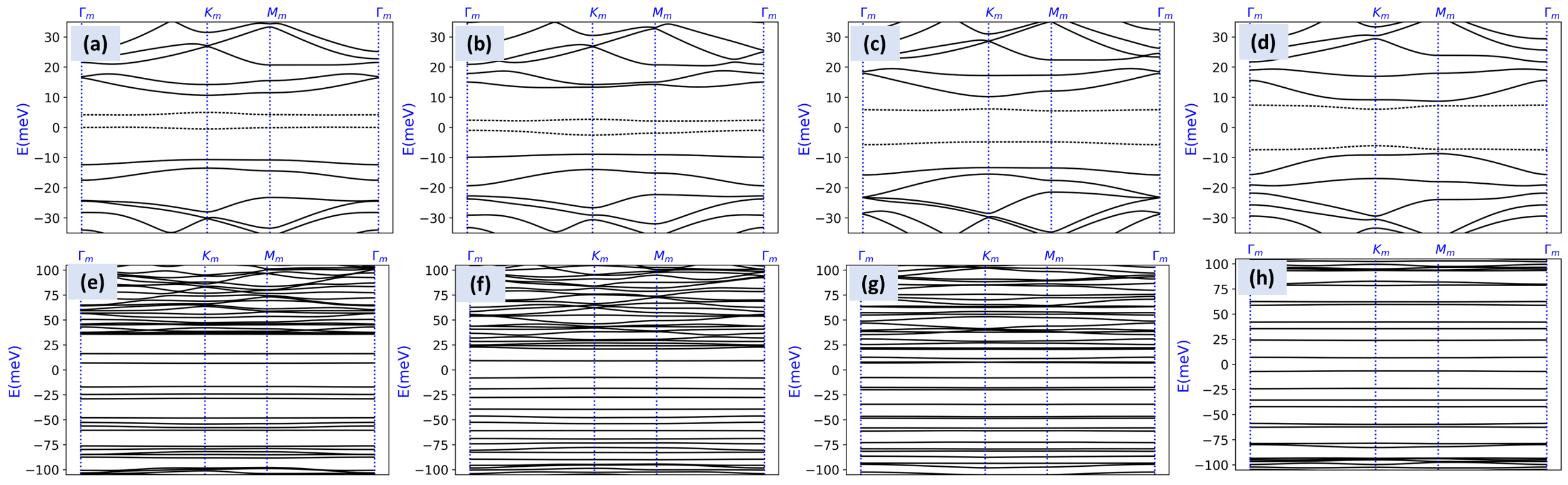}
    \caption{Evolution of the band structure of BLG in a superlattice potential versus the parameter $\alpha$ in the limit of weak (first row) and relatively strong (second row) potentials: (a,e) $\alpha=0.2$, (b,f) $\alpha=0.4$, (c,g) $\alpha=1$, (d,h) $\alpha=-1$. The potential strengths are (a) $V_{SL}=10,\,V_0=-5$ meV, (b)  $V_{SL}=10,\,V_0=-5$ meV, (c)  $V_{SL}=5,\,V_0=-10$ meV, (d)  $V_{SL}=5,\,V_0=12$ meV, (e)  $V_{SL}=40,\,V_0=-50$ meV, (f)  $V_{SL}=30,\,V_0=-50$ meV, (g)  $V_{SL}=30,\,V_0=-80$ meV, (h)  $V_{SL}=30,\,V_0=-65$ meV.      }
    \label{fig:SMvsalpha}
\end{figure}

The parameter $\alpha$ determines the ratio of the superlattice potential felt by the second layer of BLG relative to the first layer.
Here we show that the two main regimes discussed in the main text, i.e., the topological flat bands and stacks of flat bands, do not depend sensitively on the value of $\alpha$.
For small fields, topological flat bands always appear, indicated by the dashed bands in Fig.~\ref{fig:SMvsalpha}(a-d).
The stacks of flat bands also appear, but are less uniformly distributed for smaller values of $\alpha$.
Taking $\alpha \sim0.25-0.5$ yields the most uniform stacks.

As proved below and summarized in Table~\ref{tab:SymmSumm}, the Hamiltonian has special symmetry properties when $\alpha = \pm 1$; in particular, the spectrum is particle-hole symmetric when $\alpha =-1$ for any values of $V_{SL}$ and $V_0$ (summarized in Table~\ref{tab:PHSymms}).
This is reflected by the plots in Fig.~\ref{fig:SMvsalpha}(d,h). As a result, the flat bands span the conduction and valence bands symmetrically in this case.

\section{Square superlattice potential}
\begin{figure}[H]
    \centering
    \includegraphics[width=1.0\textwidth]{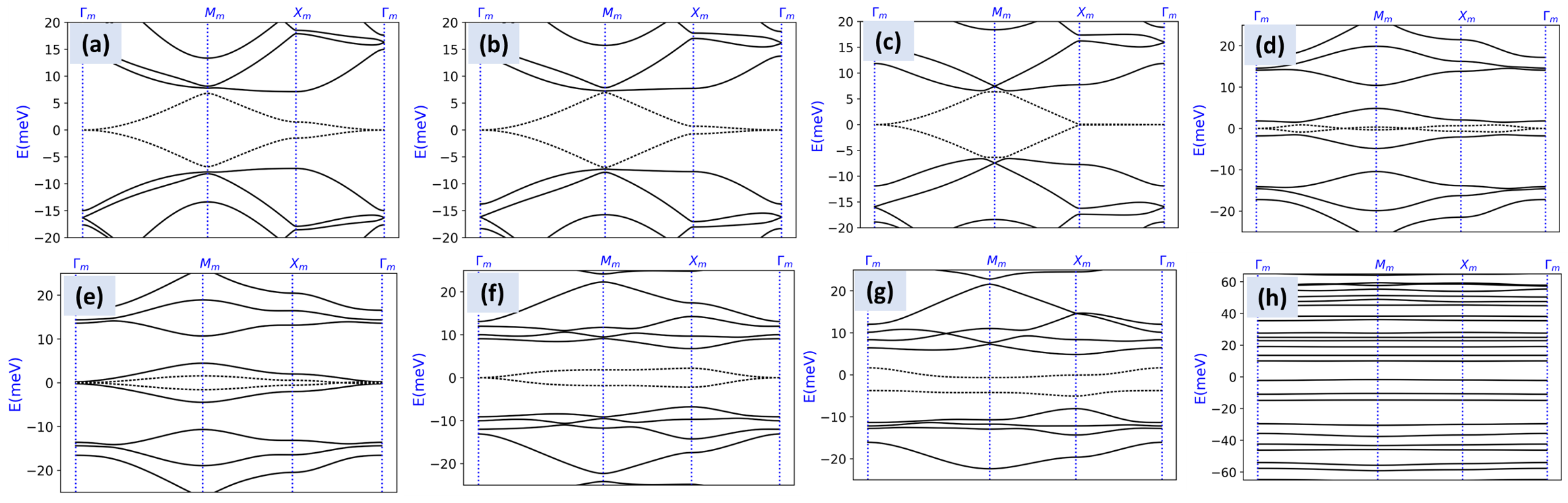}
    \caption{Evolution of the band structure of BLG in presence of a square lattice superlattice potential. (a-f) shows the spectrum with $V_0=0$ meV and $V_{SL}=5,\,7,\,9.5,\,20,\,21.5,\,\text{ and }30$ meV, respectively. In (g), $V_{SL}=30$, $V_0=-5$ meV, and in (h), $V_{SL}=45$, $V_0=-65$ meV.}
    \label{fig:SMvsSQ}
\end{figure}

To investigate the role of superlattice geometry, here we show the band structure for a square superlattice and observe qualitatively the same phenomena as on the triangular lattice, keeping the same parameters $\alpha = .3$ and $L=50$ nm.
$V_{SL}$ causes the two low-energy bands (denoted by dashed lines) to detach from the rest of the bands by opening a gap at the BZ boundary.
Turning up $V_{SL}$ causes a band inversion with higher energy bands (Fig.~\ref{fig:SMvsSQ}(b)), which changes the topology.
The critical value at which this band inversion happens is smaller than for the triangular lattice;
thus, the square lattice exhibits a smaller regime of topological flat bands in the limit of a weak superlattice potential.
Further increasing $V_{SL}$ drives several band inversions, shown in Fig.~\ref{fig:SMvsSQ}(c-f),
throughout which the two low-energy bands remain fairly well isolated energetically.
When small $V_0$ is added, a gap opens at $\Gamma_m$ to separate these two bands, forming an isolated Chern band, shown in \ref{fig:SMvsSQ}(g), where the upper dashed band carry nonzero $C$.

For larger potentials, the stack of flat bands appears. The critical value where the stack of flat bands appears is smaller than on the triangular lattice.

There are several other qualitative differences between the hexagonal and square lattices. As can be seen from Fig.~\ref{fig:SMvsSQ}(a-f) for a square superlattice, unlike the triangular lattice, the low-energy bands(dashed lines) do not gap out even after merging with other bands (Fig.~\ref{fig:SMvsSQ}(e)) and at a strong $V_{SL}$.


Finally, the spectrum for a square superlattice potential exhibits particle-hole symmetry when either $\alpha = +1$ or $V_0=0$, as summarized in Table~\ref{tab:SymmSumm}.
This extra symmetry results from the fact that the square lattice potential is symmetric under shifting by half a lattice vector and inverting sign ($V_{SL} \mapsto -V_{SL}$).

\section{Symmetries}
We now consider the pertinent symmetries of the Hamiltonian $\hat{H}$ in Eq.~(\ref{BLGSL}).
Without any applied potentials, $\hat{H}_{BLG}$ in Eq.~(\ref{HBLG}) is invariant under inversion symmetry (the inversion center is on an AB stacking point), $\hat{P}: \hat{\Psi}(\mathbf{r}) \mapsto \sigma_x \tau_x \mu_x \hat{\Psi}(-\mathbf{r})$, where $\sigma$, $\tau$, and $\mu$ are Pauli matrices that indicate sublattice, layer, and valley, as well as  time-reversal symmetry, $\hat{T}$, which is implemented by complex conjugation and also swaps the valleys.  Thus, a single valley is invariant under the combined operation $\hat{P}\hat{T}$.
Since a perpendicular displacement field is odd under inversion symmetry and even under time-reversal, $\hat{H}_{V_0}$ in Eq.~(\ref{HV0}) is odd under $\hat{P}\hat{T}$.

The superlattice potential also breaks inversion symmetry; in general $\hat{H}_{SL}$ in Eq.~(\ref{HSL}) is neither even nor odd under $\hat{P}\hat{T}$. However, for the special values of $\alpha = +1$ and $\alpha = -1$, which correspond the same(opposite) superlattice potential felt by the two layers of BLG, $\hat{H}_{SL}$ is even(odd) under inversion symmetry.


We now describe the role of continuous translations.
Since $\hat{H}_{BLG} + \hat{H}_{V_0}$ is a continuum Hamiltonian, it is also invariant under a continuous shift of the origin, i.e., under the transformation $\hat{T}_\delta : \hat{\Psi}_\mathbf{r} \mapsto \hat{\Psi}_{\mathbf{r} + \mathbf{\delta}}$.
In general, $\hat{H}_{SL}$ is not invariant under such a transformation (although its spectrum is invariant since the transformation is unitary).
However, a square superlattice potential has the special property that it is invariant under simultaneously translating by $\frac{L}{2} \hat{\mathbf{x}}$ (or $\frac{L}{2} \hat{\mathbf{y}}$) and inverting the sign of the potential, $V_{SL} \mapsto -V_{SL}$.
Thus, for the special case of a square superlattice potential, $\hat{H}_{SL}$ is odd under $\hat{T}_{1/2}^\text{sq} \equiv \hat{T}_{\frac{L}{2} \hat{\mathbf{x}}}^{\text{sq}}$, where the superscript indicates that this symmetry is specific to the square lattice.

Finally, the continuum model of bilayer graphene without any applied fields, $\hat{H}_{BLG}$, exhibits a particle-hole symmetry due to the fact that $H_{BLG}$  in Eq.~(\ref{HBLG}) is odd under $\sigma_z$. The other terms in the Hamiltonian, $H_{V_0}$ and $H_{SL}$, are even under $\sigma_z$.


\begin{table}[htb!]
\begin{tabular}{|c||c|c|c|c|c|}\hline
     & $\hat{H}_{BLG}$ & $\hat{H}_{V_0}$ & $\hat{H}_{SL}(\alpha=-1)$ & $\hat{H}_{SL}(\alpha=+1)$ & $\hat{H}_{SL}$ (general $\alpha$)\\\hline\hline
     $\sigma_z$ & $-1$ & $+1$ & $+1$ & $+1$ & $+1$\\\hline
     $\hat{P}\hat{T}$ & $+1$ & $-1$ & $-1$ & $+1$ & $X$\\\hline
     $\hat{T}_{1/2}^\text{sq}$ & $+1$ & $+1$ & $-1$ & $-1$ & $-1$\\\hline
\end{tabular}
\caption{\label{tab:SymmSumm}
Summary of (anti-)commutations relations of each term in the Hamiltonian in Eq.~(\ref{BLGSL}) with the operators in the first column.
A $+1$ indicates commutation of that operator with that part of the Hamiltonian, and a $-1$ indicates anti-commutation. X indicates neither. All operators leave a single valley invariant. In the cases where there exists an operator that anti-commutes with all terms in the Hamiltonian, the spectrum is particle-hole symmetric.}
\end{table}

The (anti-)commutation relations we have described are summarized in Table~\ref{tab:SymmSumm}.
Combining these relations shows that there are three cases where the spectrum will be particle-hole symmetric (summarized in  Table~\ref{tab:PHSymms}): first, when $\alpha=-1$, the combination $\sigma_z \hat{P}\hat{T}$ anti-commutes with the full Hamiltonian $\hat{H}$, for any geometry of the superlattice potential.
Second, when $\alpha=+1$, the combined operator $\hat{T}_{1/2}^\text{sq}\sigma_z \hat{P}\hat{T}$ anti-commutes with $\hat{H}$ in the case of a square superlattice potential.
Finally, for any $\alpha$, if there is no displacement field ($V_0 = 0$), $\sigma_z\hat{T}_{1/2}^\text{sq}$ anti-commutes with $\hat{H}$.
In all of these situations, the spectrum exhibits a particle-hole symmetry.

\begin{table}[htb!]
    \centering
    \begin{tabular}{|c||c|c|c|}\hline
         Operator & $\alpha$ & $V_0$ & Lattice \\\hline\hline
         $\sigma_z\hat P\hat T$ & $-1$ & Any & Any\\\hline
         $\hat{T}_{1/2}^\text{sq}\sigma_z \hat{P}\hat{T}$ & $+1$ & Any & Square\\\hline
         $\sigma_z\hat{T}_{1/2}^\text{sq}$ & Any & $0$ & Square\\\hline
    \end{tabular}
    \caption{Operators that can yield a particle-hole symmetry of the spectrum and the conditions under which they anti-commute with the Hamiltonian.}
    \label{tab:PHSymms}
\end{table}

\section{Effect of the relative sign of $V_0$ and $V_{SL}$}

Here we provide an example showing that when $V_0$ and $V_{SL}$ have opposite signs, a weaker $V_0$ is required to realize the stack of flat bands.

\begin{figure}[H]
    \centering
    \includegraphics[width=0.8\textwidth]{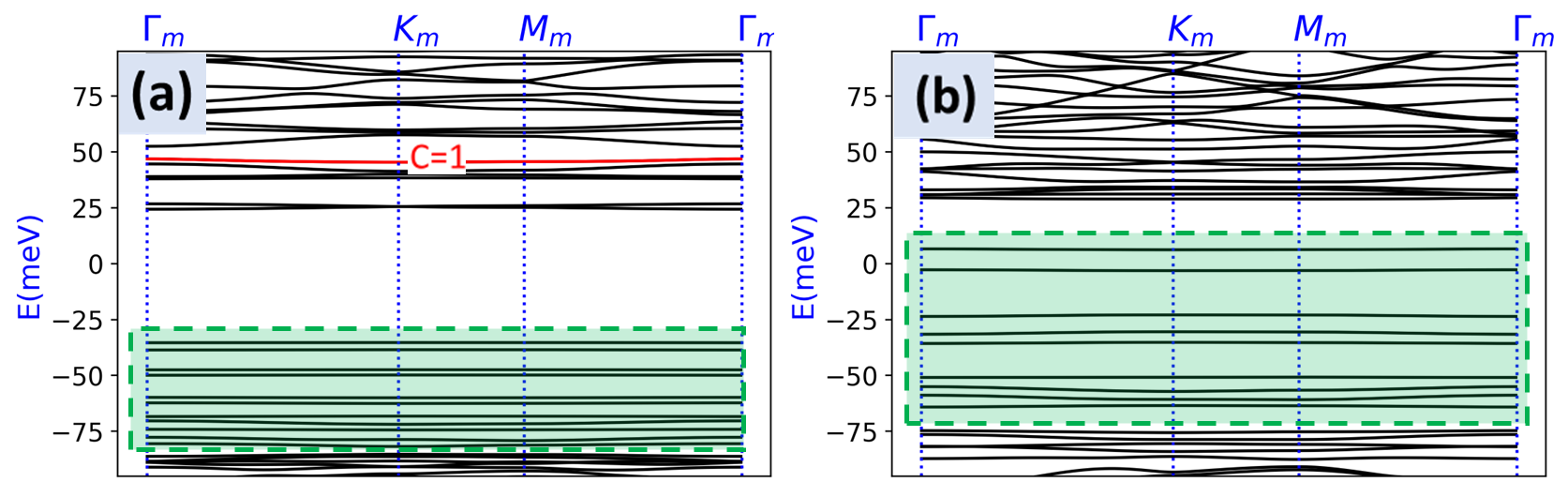}
    \caption{Band structures for intermediate $V_{SL}$, $V_0$. Band structure of Eq.~(1) in the main text with (a) $V_{SL}=24,\,V_0=80$ meV and (b) $V_{SL}=24,\,V_0=-45$ meV. Red lines show topological bands with their corresponding Chern numbers. Shaded green area shows the stack of flat bands.}
    \label{fig:V0_nonzero}
\end{figure}

\section{Trigonal warping}
Here we show the analogue of Fig. 2 of the main text with trigonal warping included. We choose $v_w= 0.1 v$ as the trigonal correction to the Fermi velocity, implemented by:
\begin{align}\label{warping}
   H_{w}(\vex{k})=\hbar v_w \left(
  \begin{array}{cccc}
    0 & 0 & 0 & \chi k_x - i k_y\\
    0 & 0 & 0 & 0 \\
    0 & 0 & 0 & 0 \\
    \chi k_x + i k_y & 0 & 0 & 0
  \end{array}
\right)
\end{align}

\begin{figure}[H]
    \centering
    \includegraphics[width=0.8\textwidth]{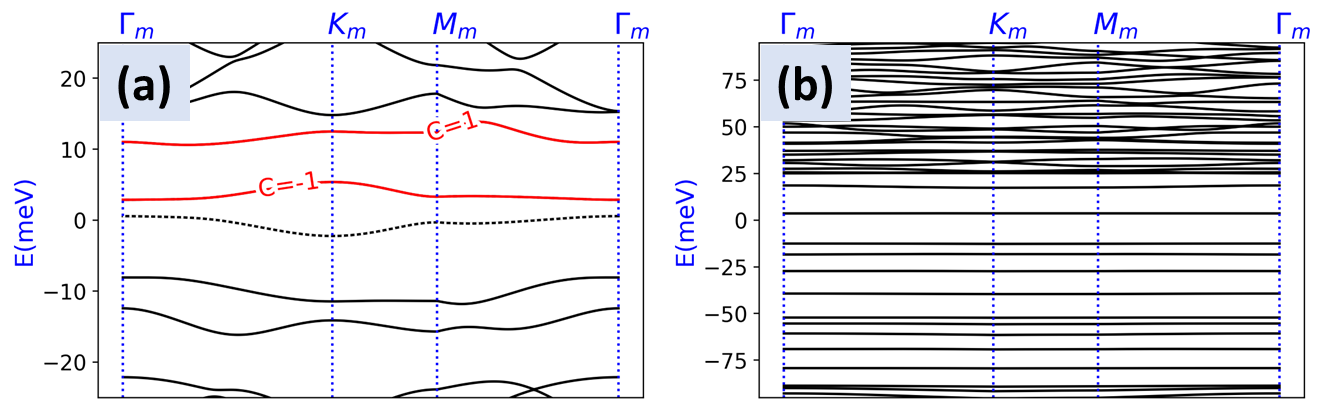}
    \caption{Band structures with trigonal warping. The parameters are the same as in Fig. 2 of the main text but include the triagonal warping term in Eq.~(\ref{warping}).}
    \label{fig:triwarping}
\end{figure}

Comparing Fig.~\ref{fig:triwarping} to Fig. 2 in the main text shows that the effect of trigonal warping is minimal.
However, trigonal warping can cause band inversions between bands whose separation is less than the scale of the trigonal warping term.
For example, in the small field limit, the trigonal warping term induces a topological phase transition where the second lowest conduction band goes from $C=-2$ in Fig 2(a) of the main text to $C=1$ in Fig.~\ref{fig:triwarping}(a).
The effect of trigonal warping in the stacked band regime is negligible, as shown by comparing Fig.~\ref{fig:triwarping}(b) to Fig. 2(b) in the main text.
Overall, the effect of trigonal warping in our calculation can be compensated for by tuning other knobs in the Hamiltonian.

\section{Many-body calculation of a FCI}

To address the possibility of fractional Chern insulator (FCI) states upon partial filling of the Chern $ |C|  = 1 $ band in the limit of weak superlattice potential,
we study the many-body Hamiltonian projected into the flat band.

\subsection*{Projection onto the Chern $|C| = 1$ band}

We first describe the relevant projected operators.
The density operator in momentum space around one valley for a single spin is defined as 	
	\begin{equation}
		\label{eq:denoperator}
		\rho_{\tau\sigma} (\mathbf{q}) = \sum_{\mathbf{k}} \hat{\Psi}^\dagger_{\tau\sigma}(\mathbf{k} + \mathbf{q}) \hat{\Psi}_{\tau\sigma}(\mathbf{k}),
	\end{equation} where $\tau$ and $\sigma$ are indices that label the valley and spin degrees of freedom respectively.
	$\hat{\Psi}(\mathbf{k})$ is the electron annihlation operator with momentum $\mathbf{k}$.
	By diagonalizing the single-particle Hamiltonian, $\hat{\Psi}(\mathbf{k})$  can be written in terms of the band basis as
\begin{equation}
	\hat{\Psi}_{\tau \sigma}(\mathbf{k} + m \mathbf{G}_1 + n \mathbf{G}_2) = \sum_{\alpha} u^{mn}_{\alpha \tau \sigma}(\mathbf{k}) \hat{c}^\dagger_{\alpha \tau \sigma}(\mathbf{k}),
\end{equation}
	where $\alpha$ runs over all bands and $ u^{mn}_{\alpha \tau \sigma}(\mathbf{k})$ is the respective component of the band eigenstate plane wave expansion in terms of the two reciprocal superlattice vectors $\mathbf{G}_1$ and $\mathbf{G}_2$. $\hat{c}^\dagger_{\alpha \tau \sigma}(\mathbf{k}) $ is the band creation operator. In the following, we use a periodic gauge by setting $u^{mn}_{\alpha \tau \sigma}(\mathbf{k} + m_0\mathbf{G}_1 + n_0\mathbf{G}_2) = u^{m + m_0,n + n_0}_{\alpha \tau \sigma}(\mathbf{k}) $.  The density operator \eqref{eq:denoperator} in the band basis becomes \begin{equation}
		\label{eq:denoperator_band}
		\rho_{\tau\sigma} (\mathbf{q}) = \sum_{\mathbf{k} \in \textrm{mBZ}, \alpha_1, \alpha_2} \lambda^{\alpha_1\alpha_2}_{\tau \sigma}(\mathbf{k} + \mathbf{q},\mathbf{k}) \hat{c}^\dagger_{\alpha_1 \tau \sigma}(P(\mathbf{k}+\mathbf{q})) \hat{c}_{\alpha_2 \tau \sigma}(\mathbf{k}),
	\end{equation}
with form factors $ \lambda^{\alpha_1\alpha_2}_{\tau \sigma}(\mathbf{k} + \mathbf{q},\mathbf{k}) = \sum_{mn} u_{\alpha_1\tau\sigma}^{* m+m_0,n+n_0}(P(\mathbf{k} + \mathbf{q}))u_{\alpha_2 \tau \sigma}^{mn}(\mathbf{k})$ for $\mathbf{k}+\mathbf{q} = \mathbf{k}_0 + m_0 \mathbf{G}_1 + n_0\mathbf{G}_2$ where $\mathbf{k}_0 \in \textrm{mBZ}$ and and $m_0,n_0$ are two integers.
$P(\mathbf{k})$ is a projector to the mini Brillouin zone (mBZ) , i.e, $ P(\mathbf{k}) = \mathbf{k}_0$ for $\mathbf{k} = \mathbf{k}_0 + m_0\mathbf{G}_1 + n_0 \mathbf{G}_2$. The projection to a band (or a set of bands $\{\beta\}$) is obtained by truncating the sum over $\alpha$ in \eqref{eq:denoperator_band} to include only the relevant bands. This defines the projected density operator, \begin{equation}
\tilde{\rho}_{\tau\sigma} (\mathbf{q}) = \sum_{\mathbf{k} \in \textrm{mBZ}, \beta_1, \beta_2} \lambda^{\beta_1\beta_2}_{ \tau \sigma}(\mathbf{k} + \mathbf{q},\mathbf{k}) \hat{c}^\dagger_{\beta_1 \tau \sigma}(P(\mathbf{k}+\mathbf{q})) \hat{c}_{\beta_2 \tau \sigma}(\mathbf{k}).
\end{equation}
In the following, we take $\beta$ to denote the Chern $|C| = 1$ band in the limit of weak superlattice potential, i.e., the band shown in Fig. 1(c) in the main text.

\subsection*{Interaction Hamiltonian}
Next, we consider density-density interactions projected onto the Chern band. They are of the form \begin{equation}
	\label{eq:intHam}
H_{\textrm{int}} = \dfrac{1}{N_c}\sum_{\mathbf{q} \in R^2, \tau_1 \tau_2 \sigma_1 \sigma_2} V(\mathbf{q}) : \tilde{\rho}_{\tau_1 \sigma_1}(\mathbf{q}) \tilde{\rho}_{\tau_2 \sigma_2}(-\mathbf{q}):,
\end{equation}
where we take the interaction to be the dual-gated Coulomb potential $V(\mathbf{q}) = 2 \pi V_C \tanh(d_g |\mathbf{q}|)/(\sqrt{3}|\mathbf{q}|L)$ with $V_C$ the strength of the interaction,
$d_g$ is the screening length, which we take to be $d_g = 50 \> \textrm{nm}$ and $L$ the periodicity of the superlattice potential. $N_c$ is the number of unit cells. The Hamiltonian \eqref{eq:intHam} has a $U(2) \times U(2)$ symmetry corresponding to independent charge and spin conservation \textit{within} each valley. Strictly speaking, the full density operator also contains an intervalley contribution in addition to the intravalley term \eqref{eq:denoperator} that we do not include here. Including these terms will break the $U(2) \times U(2)$ symmetry to $U_c(1) \times U_v(1) \times SU(2)$ corresponding to conservation of charge, valley and total spin. However, these are Umklapp terms that scatter between valleys  and they are suppressed by the Coulomb interaction relative to the intravalley terms as $1/L$. As a result, we neglect them in our analysis due to the enlarged periodicity of the system.


\begin{figure}
    \centering
    \includegraphics[width=\textwidth]{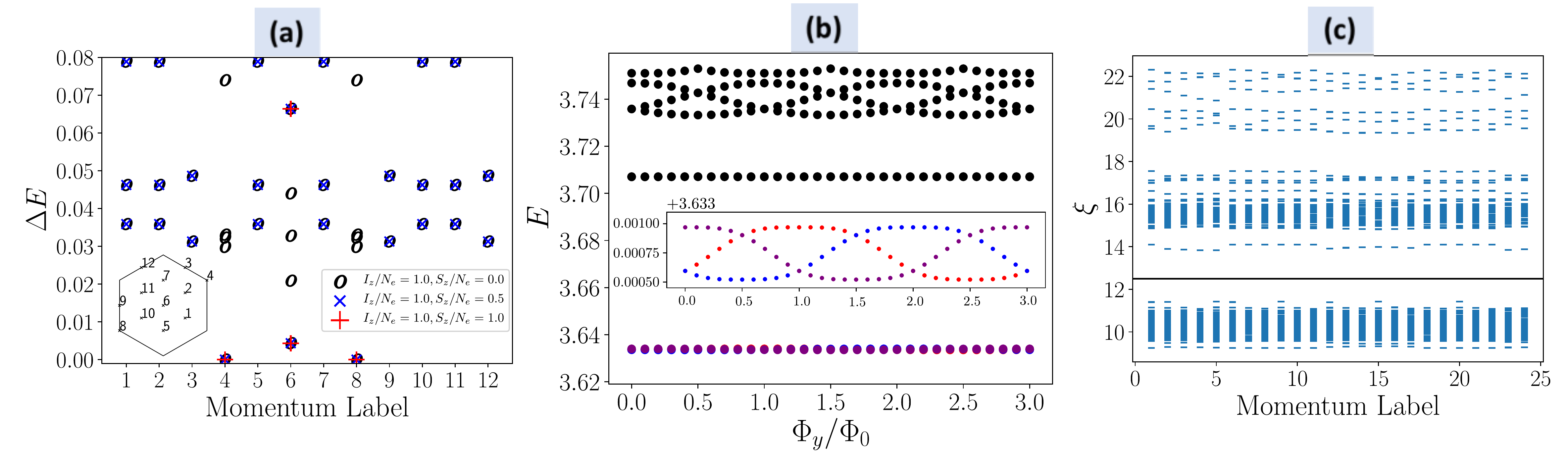}
   \caption{(a) The many-body spectrum of the Hamiltonian \eqref{eq:intHam} obtained from exact diagonalization on the finite lattice shown in the inset for $N_e = 4$ electrons. The many-body states are labelled by their total valley and spin polarization. (b) Spectral flow of the many-body spectrum of the spin and valley polarized Hamiltonian \eqref{eq:intHam} for $N_e = 8$ electrons on a finite lattice of $N_x \times N_y = 24$ unit cells, upon inserting a flux into one handle of the torus. The inset shows the spectral flow of the lowest three quasi-degenerate many-body states. (c) Particle entanglment spectrum (PES) of the three degenerate ground states after tracing out $N_B = 5$ electrons. There are 1088 states below the line corresponding to the expected number of quasi-hole excitations of Laughlin state on the finite lattice used.}
\label{fig:FCI_SM}
\end{figure}
\subsection*{Evidence of a fractional Chern insulator in the  $|C| = 1$ band}
Here, we provide evidence for a Laughlin-like FCI at partial filling $\nu = 1/3$ of the Chern $|C| = 1$ band above charge neutrality in the weak superlattice potential regime. We take the flat band limit, neglecting the small band dispersion. We then set $V_C = 1$ and exactly diagonalize the Hamiltonian \eqref{eq:intHam} on finite lattices with periodic boundary conditions, focusing on filling $\nu = 1/3$. We consider first the full problem including spin and valley degrees of freedom. In such a case, the many-body states are decomposed into different sectors that are labelled by total valley polarization $I_z = \frac{1}{N_e} \sum_{\sigma}  N_{+\sigma} - N_{-\sigma}$ and total spin polarization $S_z = \frac{1}{N_e} \sum_{\tau}  N_{\tau \uparrow} - N_{\tau \downarrow}$, where $N_{\tau \sigma}$ is the number of electrons in valley $\tau$ carrying spin $\sigma$. Fig. \ref{fig:FCI_SM}(a) shows that the three lowest energies correspond to states that are valley polarized ($I_z = 1$), in addition to being degenerate in all three spin sectors $S_z$. This indicates both maximal $U(1)$ valley polarization and maximal $SU(2)$ spin polarization.

Having established spin and valley polarization on a small system, we perform exact diagonalization on a bigger system where we assume spin and valley polarization.
A promising sign for an FCI is the expected threefold ground state degeneracy on the torus, shown in Fig.~4(b) in the main text.
However, this does not rule out the existence of competing states such as charge density waves (CDWs) which would show the same degeneracy. To confirm the FCI state, we have calculated the spectral flow of the many-body ground states upon inserting flux through one handle of the torus, shown in Fig.~\ref{fig:FCI_SM}(b).
We observe that the three degenerate ground states flow and mix into each other, and are separated by a large gap to the excited states.
After inserting three units of flux quantum, the system returns to its initial configuration, suggesting quantized fractional Hall conductance.

    In addition to the spectral flow, very compelling evidence comes from the particle entanglement spectrum (PES) calculated by partitioning the system into two parts with $N_A$ and $ N_B = N_e - N_A$ electrons then tracing out $N_B$ electrons from the system. The PES probes the nature of quasi-hole excitations of the underlying phase and can also be used to distinguish FCIs from other phases such as CDWs \cite{li_entanglement_2008,bernevig_emergent_2012,sterdyniak_extracting_2011}. As shown in Fig \ref{fig:FCI_SM}(c), we label each PES level by the total momentum, which remains a good quantum number after partitioning.  We observe a low lying sector separated by a gap to the rest of the spectrum. The number of states in this low lying sector matches the expected number of quasi-hole excitations in the abelian Laughlin universality class at filling $\nu = 1/3$ \cite{wu_haldane_2014,regnault_entanglement_2015}.

\end{document}